# On The Estimation of the Hurst Exponent Using Adjusted Rescaled Range Analysis, Detrended Fluctuation Analysis and Variance Time Plot: A Case of Exponential Distribution

Roel F. Ceballos[1] & Fe F. Largo[2]

[1]Instructor, Dept. of Mathematics and Statistics
College of Arts and Sciences
University of Southeastern Philippines, Obrero Davao City, 8000

[2]Assistant Professor, Dept. of Mathematics and Statistics
College of Arts and Sciences
University of Southeastern Philippines, Obrero Davao City, 8000

*Abstract*: Hurst Exponent has been widely used in different fields as a measure of long range dependence in time series. It has been studied in hydrology and geophysics, economics and finance, and recently, it is still a hot topic in the different areas of research involving DNA sequences, cardiac dynamics, internet traffic, meteorology and geology. Various methods in the estimation of Hurst Exponent have been proposed such as Adjusted Rescaled Range Analysis, Detrended Fluctuation Analysis and Variance Time Plot Analysis.

This study explored the efficiency of the three methods: Adjusted Rescaled Range Analysis, Detrended Fluctuation Analysis and Variance Time Plot Analysis in the estimation of Hurst Exponent when data are generated from an exponential distribution. In addition, the efficiency of the three methods was compared in different sample sizes of 128, 256, 512, 1024 and varying $\lambda$ parameter values of 0.1, 0.5, 1.5, 3.0, 5.0 and 7.0. The estimation process for each of the methods using different sample sizes and $\lambda$ parameter values were repeated for 100, 500 and 1000 times to verify the consistency of the result. A Scilab Program containing different functions was developed for the study to aid in the simulation process and calculation.

The Adjusted Rescaled Range Analysis was the most efficient method with the smallest Mean Square Error for all $\lambda$ parameter values and different sample sizes.

## 1 Introduction

Hurst Exponent (denoted by H) is a classical self-similarity parameter that measures the long-range dependence in a time series and provides measure of long-term nonlinearity[19]. The early development of Hurst exponent is traced back in 1951, when a British hydrologist, Dr. Harold Edwin Hurst, together with his team, conducted a study to determine the optimum dam sizing for the Nile River's volatile rain and drought conditions observed over a long period of time. Over the years, Hurst exponent has been applied in a wide range of industries. For example the Hurst exponent is paired with technical indicators to make decisions about trading securities in financial markets; and it is used extensively in the healthcare industry, where it is paired with machine-learning techniques to monitor EEG signals. The Hurst exponent can even be applied in ecology, where it is used to model increase and decrease in populations [18].

In this paper, the researcher decided to use the Adjusted Rescaled Range (R/Sal) Analysis, Detrended Fluctuation Analysis and Variance time Plot Analysis in the estimation of Hurst exponent in a time series data generated randomly from an exponential distribution. In section 2 we discussed the three methods and presented the estimation results and later, in section 3 we present the comparison of the three methods. The Rescaled Range Analysis (R/Sal) turns out to be the most efficient method for the estimation of Hurst Exponent considering the different sample sizes of 128, 256, 512 and 1024.

## 2 Methods for estimating Hurst Exponent

### 2.1 Adjusted Rescaled Range Analysis

The R/S analysis was originally created by the British hydrologist Harold Edwin Hurst while he was studying the problem of water storage on the Nile River. Later, it was popularized by Benoit Mandelbrot especially in the area of long term dependency analysis of the stock market. The computation for the estimation of Hurst exponent using the Adjusted Rescaled Range (R/Sal) Analysis is adopted from the work of Weron[25]. A detailed step by step procedure is herein presented: Divide the time series $Z$ of length $N$ into $d$ subseries of length $n$, where $n$ is an integral divisor of $N$. For each subseries $m = 1, 2, \ldots, d$, the following steps will be followed.

Step 1. Find the mean $E_m$ and Standard deviation $S_m$.
Step 2. Normalize the data $(Z_{i,m})$ by subtracting the sample mean
$$X_{i,m} = Z_{i,m} - E_m \quad for\ i = 1, 2, \ldots, n \tag{1}$$
Step 3. Create a cumulative time series
$$Y_{i,m} = \sum_{j=1}^{i} X_{j,m} \quad for\ i = 1, 2, \ldots, n \tag{2}$$
Step 4. Find the Range
$$R_m = max(Y_{1,m}, \ldots, Y_{n,m}) - min(Y_{1,m}, \ldots, Y_{n,m}) \tag{3}$$
Step 5. Rescale the range by dividing it with the standard deviation, $R_m/S_m$.
Step 6. Calculate for the mean value of the rescaled range for all subseries of length $n$ using the formula
$$(R/S)_n = (1/d)(\sum_{m=1}^{d} R_m/S_m) \tag{4}$$
Step 7. The R/S statistics follows the relation $(R/S)_n \sim cn^H$. Thus, the value of $H$ can be obtained by running a simple linear regression with $log(n)$ as the independent variable and $log(R/S)_n$ as the dependent variable. The slope of the resulting equation in (5) of an ordinary least squares regression is the estimate of the Hurst exponent.
$$log(R/S)_n = log c + H log(n) \tag{5}$$
However, for small values of $n$, there is a significant deviation from the 0.5 slope. For this reason, the theoretical values for *R/S* Statistics are usually approximated by:

$$E(R/S)_n = \begin{cases} \left(\frac{n-1/2}{n}\right)\left(\frac{\Gamma(n-1/2)}{\sqrt{\pi}\Gamma(n/2)}\right)\left(\sum_{i=1}^{n-1}\sqrt{\frac{n-i}{i}}\right) & \text{for } n \leq 340 \\ \left(\frac{n-1/2}{n}\right)\left(1/\sqrt{\frac{n\pi}{2}}\right)\left(\sum_{i=1}^{n-1}\sqrt{\frac{n-i}{i}}\right) & \text{for } n > 340 \end{cases} \tag{6}$$

where $\Gamma$ is the Euler gamma function. This formula is a slight modification of the formula given by Anis and Lloyd (1976); the $\left(\frac{n-1/2}{n}\right)$ was added by Peters (1994) to improve the performance for very small $n$. Then the adjusted Rescaled Range statistics will be
$$(R/Sal)_n = (R/S)_n - E(R/S)_n + E(R/Sal)_n \tag{7}$$
where $E(R/Sal)_n = \sqrt{0.5\pi n}$

The estimate for the value of Hurst exponent using adjusted Rescaled Range Statistics will be the slope obtained by running a simple linear regression with $log(R/Sal)_n$ as dependent variable and $log n$ as independent variable.

The estimates of Hurst Exponents and the Mean Square Error obtained using the Adjusted Rescaled Range (RSal) Analysis are shown in Table 1. The estimates of Hurst Exponent get close to 0.5 as the sample size and iterations becomes large. The Mean Square Errors become smaller as the sample size become larger. These observations are generally noticeable for the Hurst estimates and MSE in different λ parameter values.

**Table 1. Hurst Estimates and MSE using Adjusted Rescaled Range (RSal)**

| | N= | 128 | 256 | 512 | 1024 | 128 | 256 | 512 | 1024 |
|---|---|---|---|---|---|---|---|---|---|
| λ | Iteration | \multicolumn{4}{c}{Average Hurst Estimates} | | | | Mean Square Error | | | |
| | 100 | 0.4964 | 0.4994 | 0.4982 | 0.4963 | 0.0014 | 0.0008 | 0.0006 | **0.0005** |
| **0.1** | 500 | 0.5003 | 0.4999 | 0.4985 | 0.4994 | 0.0012 | 0.0010 | 0.0006 | **0.0005** |
| | 1000 | 0.4986 | 0.4992 | 0.4991 | 0.4994 | 0.0015 | 0.0009 | 0.0006 | **0.0005** |
| | 100 | 0.5057 | 0.5020 | 0.502 | 0.4989 | 0.0012 | 0.0009 | 0.0006 | **0.0004** |
| **0.5** | 500 | 0.4984 | 0.4994 | 0.5005 | 0.4984 | 0.0014 | 0.0010 | 0.0007 | **0.0005** |
| | 1000 | 0.4999 | 0.4992 | 0.4987 | 0.4988 | 0.0013 | 0.0009 | 0.0006 | **0.0005** |
| **1.5** | **100** | 0.5013 | 0.502 | 0.5025 | 0.4994 | 0.0013 | 0.0009 | 0.0005 | **0.0005** |
|  | **500** | 0.4988 | 0.4983 | 0.4982 | 0.4991 | 0.0015 | 0.0010 | 0.0007 | **0.0005** |
|  | **1000** | 0.4986 | 0.4982 | 0.4986 | 0.4984 | 0.0015 | 0.0009 | 0.0007 | **0.0005** |
| **3.0** | **100** | 0.502 | 0.4973 | 0.4969 | 0.5003 | 0.0014 | 0.0009 | 0.0005 | **0.0005** |
|  | **500** | 0.4989 | 0.5005 | 0.4989 | 0.4996 | 0.0013 | 0.0009 | 0.0007 | **0.0004** |
|  | **1000** | 0.4998 | 0.500 | 0.4988 | 0.4994 | 0.0013 | 0.0010 | 0.0007 | **0.0005** |
| **5.0** | **100** | 0.5016 | 0.5017 | 0.4983 | 0.4982 | 0.0017 | 0.0009 | 0.0008 | **0.0005** |
|  | **500** | 0.5007 | 0.4974 | 0.499 | 0.4999 | 0.0013 | 0.0009 | 0.0006 | **0.0005** |
|  | **1000** | 0.5027 | 0.4986 | 0.4993 | 0.4995 | 0.0014 | 0.0009 | 0.0006 | **0.0005** |
| **7.0** | **100** | 0.4965 | 0.5001 | 0.4959 | 0.4953 | 0.0017 | 0.0009 | 0.0006 | **0.0004** |
|  | **500** | 0.5030 | 0.4974 | 0.5007 | 0.499 | 0.0015 | 0.0009 | 0.0007 | **0.0005** |
|  | **1000** | 0.5012 | 0.5007 | 0.4993 | 0.4994 | 0.0014 | 0.0009 | 0.0006 | **0.0005** |

The graphical presentations of the average estimates of the Hurst Exponent are shown in Figure 1. Three graphs are presented, one for every iteration of 100, 500 and 1000. The observations discussed in Table 1 can be seen clearly by looking at the graphs. The estimates of Hurst Exponent at N=1024 are the closest estimates to 0.5 for all λ parameter values. Apparently, an increase in the sample size improves the values of the estimates and their corresponding Mean Square Errors.

**Figure 1. Plot of the Average Hurst Estimates using R/Sal**

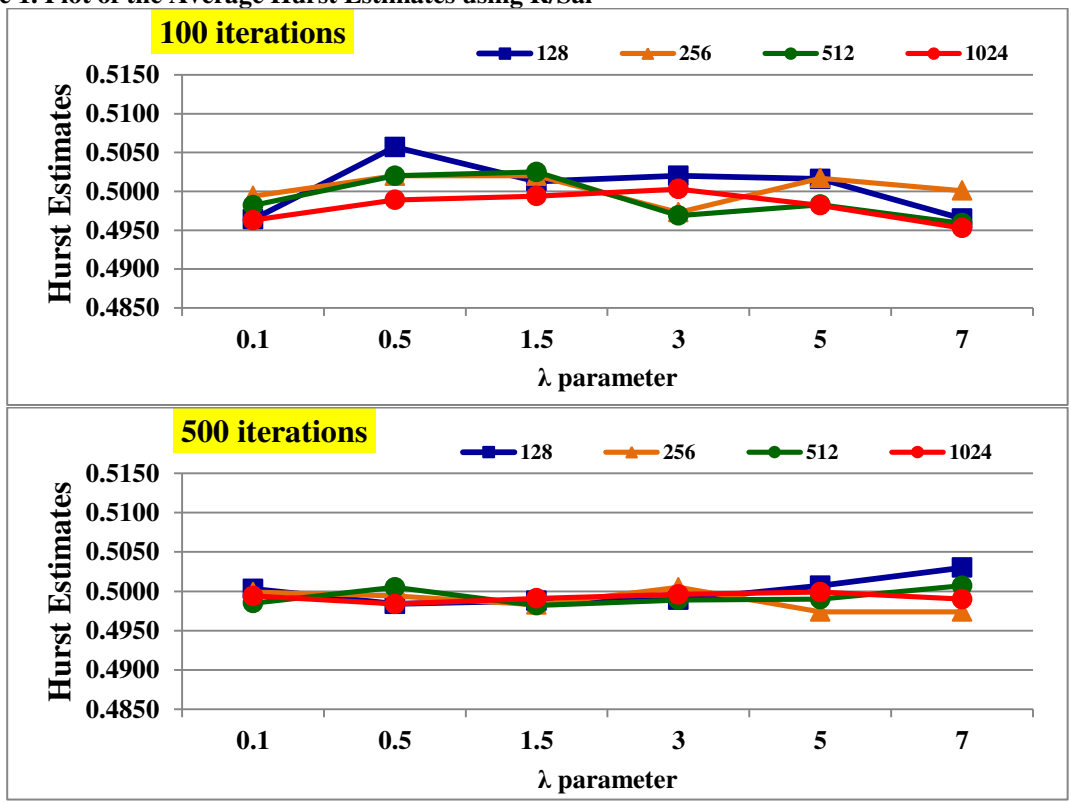

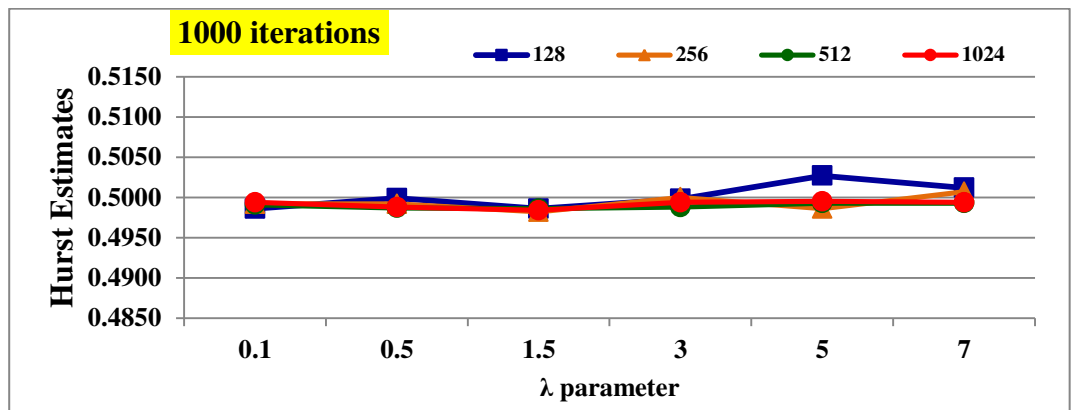

## 2.2 Detrended Fluctuation Analysis

Detrended Fluctuation Analysis is a method proposed by Peng et al. (1994) in a paper that has been cited more than 2000 times and represents an extension of the ordinary fluctuation analysis. It is used to measure long range dependence in a time series. While the significant Hurst Exponent value is between 0 and 1, it is possible for DFA to produce Hurst Exponent values greater than 1. Hurst values greater than 1 indicate non-stationarity or unsuccessful detrending (Bryce et al., 2001). The computational procedure for the Detrended Fluctuation Analysis in this study is adopted from the paper presented by Weron (2001). The procedures of the method can be summarized as follows: Divide the time series $X$ of length $N$ into $d$ subseries of length $n$. Next for each subseries, $m = 1, \dots, d$, the following steps will be executed.

Step 1. Create a cumulative time series
$$Y_{i,m} = \sum_{j=1}^{i} X_{j,m} \quad for \quad i = 1, \dots, n \qquad (8)$$
Step 2. Fit a least squares line
$$\hat{Y}_m(t) = a_m * t + b_m \; to\{Y_{1,m}, Y_{2,m}, \dots, Y_{n,m}\} \qquad (9)$$
where $t = 1, \dots, n$
Step 3. Calculate the root mean square fluctuation of the detrended time series
$$F(m) = \sqrt{\frac{1}{n}\sum_{i=1}^{n}(Y_{i,m} - a_m(i) - b_m)^2} \qquad (10)$$
Step 4. Calculate the mean value of the root mean square for all subseries of length n.
$$\mathcal{F}(n) = \frac{1}{d}\sum_{m=1}^{d} F(m) \qquad (11)$$
Step 5. Finally, just like the R/S method, the value of H can be obtained by running a simple linear regression with $\log n$ as the independent variable and $\log F(n)$ as the dependent variable.

Presented in Table 2 are the estimates of the Hurst Exponents and the Mean Square Error obtained using the Detrended Fluctuation Analysis (DFA). Although the estimates of Hurst Exponents are above 0.50, there are improvements in the estimates when the sample size is increased. The estimates of Hurst Exponent are lowest and closest to 0.50 when *N*=1024 and with 1000 iterations for all λ parameter values. Consequently, the values of the Mean Square Error are smallest when *N*=1024. Hence, the estimate of Hurst Exponent is most efficient when the sample size N is increased.

**Table 3. Hurst Estimates and MSE using Detrended Fluctuation Analysis**

| λ | N= Iteration | 128 | 256 | 512 | 1024 | 128 | 256 | 512 | 1024 |
|---|---|---|---|---|---|---|---|---|---|
|   |   | Average Hurst Estimates | | | | Mean Square Error | | | |
| 0.1 | 100 | 0.5916 | 0.5817 | 0.5559 | 0.5427 | 0.0152 | 0.0097 | 0.005 | **0.0031** |
|     | 500 | 0.6024 | 0.5761 | 0.5590 | 0.5474 | 0.0161 | 0.0091 | 0.0054 | **0.0035** |
|     | 1000 | 0.5957 | 0.5761 | 0.5597 | 0.5478 | 0.0155 | 0.0091 | 0.0054 | **0.0035** |
| 0.5 | 100 | 0.6063 | 0.5711 | 0.5681 | 0.5481 | 0.0172 | 0.0075 | 0.0067 | **0.0034** |
|     | 500 | 0.5954 | 0.5798 | 0.5621 | 0.5470 | 0.0155 | 0.0098 | 0.0058 | **0.0035** |
|     | 1000 | 0.5997 | 0.5772 | 0.5588 | 0.5473 | 0.0162 | 0.0094 | 0.0054 | **0.0036** |
| 1.5 | 100 | 0.6009 | 0.5828 | 0.5603 | 0.5486 | 0.0162 | 0.0105 | 0.0053 | **0.0035** |
|     | 500 | 0.5959 | 0.5767 | 0.5562 | 0.5474 | 0.0155 | 0.0093 | 0.0052 | **0.0034** |
|     | 1000 | 0.5957 | 0.5741 | 0.5607 | 0.5469 | 0.0155 | 0.0087 | 0.0058 | **0.0035** |
| 3.0 | 100 | 0.6079 | 0.5757 | 0.5552 | 0.5457 | 0.0172 | 0.0089 | 0.0047 | **0.0034** |
|     | 500  | 0.6013 | 0.5760 | 0.5596 | 0.5482 | 0.0163 | 0.0092 | 0.0055 | **0.0035** |
|     | 1000 | 0.5997 | 0.5764 | 0.5588 | 0.5478 | 0.0162 | 0.0092 | 0.0053 | **0.0035** |
|     | 100  | 0.6046 | 0.5813 | 0.5590 | 0.5455 | 0.0180 | 0.0101 | 0.0056 | **0.0034** |
| 5.0 | 500  | 0.5980 | 0.5777 | 0.5575 | 0.5485 | 0.0161 | 0.0092 | 0.0052 | **0.0037** |
|     | 1000 | 0.6053 | 0.5779 | 0.5603 | 0.5479 | 0.0174 | 0.0093 | 0.0056 | **0.0035** |
|     | 100  | 0.5980 | 0.5746 | 0.5534 | 0.5419 | 0.0166 | 0.0092 | 0.0044 | **0.0028** |
| 7.0 | 500  | 0.6075 | 0.5711 | 0.5630 | 0.5474 | 0.0181 | 0.0083 | 0.0059 | **0.0036** |
|     | 1000 | 0.6038 | 0.5792 | 0.5603 | 0.5478 | 0.0174 | 0.0095 | 0.0056 | **0.0035** |

The graphical presentations of the average estimates of the Hurst Exponent using Detrended Fluctuation Analysis (DFA) are shown in Figure 2. There are three graphs, one for each of the following iterations; 100, 500 and 1000. The graph shows that the estimates are getting close to 0.50 as the *N* is increased.

**Figure 2. Plot of the Average Hurst Estimates using DFA**

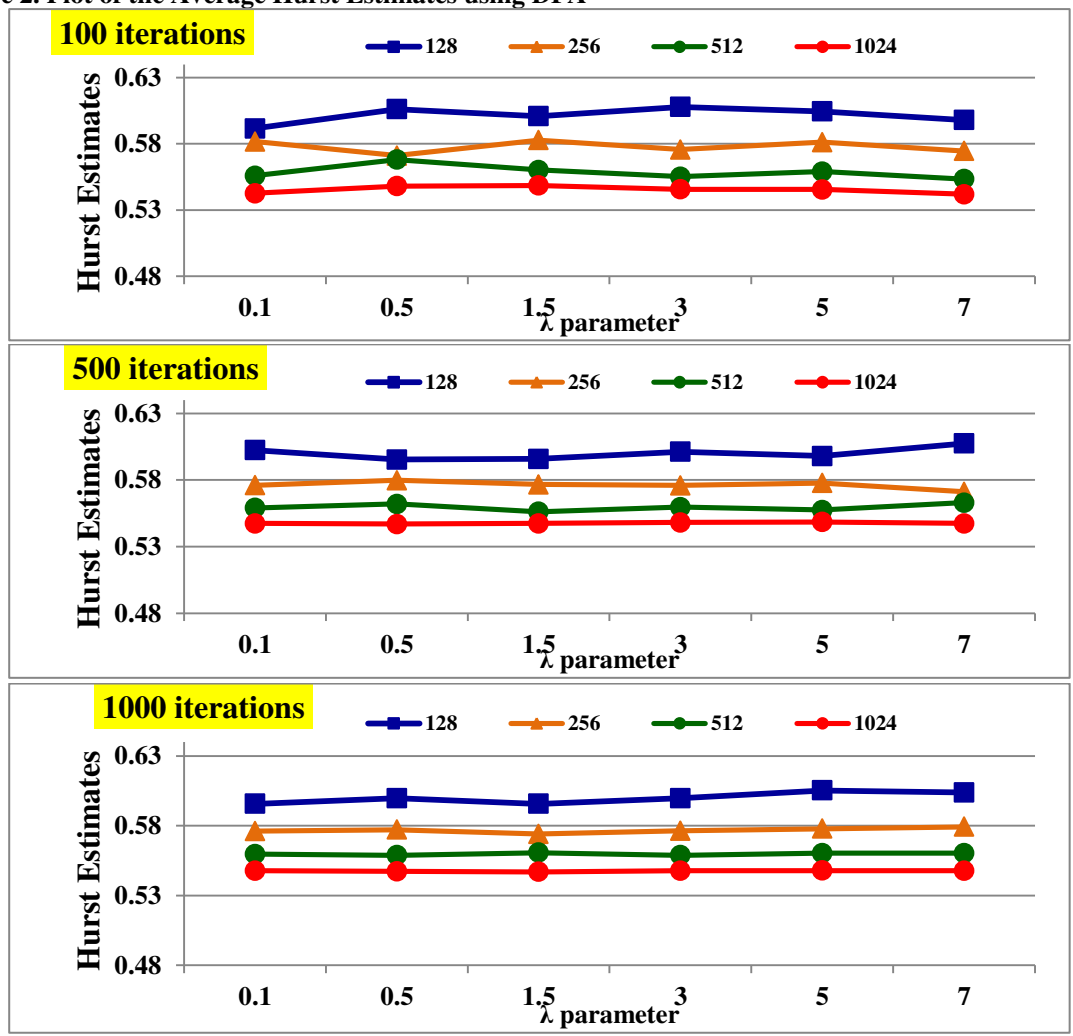

### 2.3 Variance –Time Plot Analysis

The variance time plot analysis is based on property of slow decaying variance of self-similar processes undergoing aggregation. Pujolle et al. (2000) said it is based on the asymptotic relationship of the variance of sample averages $X_m$ of non-overlapping blocks of data of size m from the process *X*, with the relationship given by:

$$Var(X^{(w)}) \sim cw^{-\beta}, as\ w \to \infty$$

The method is widely used to analyze telecommunications data and real network traffic data. This method is quite different from the first two since variance time will not use the integral divisors of *N* to create subseries or sub windows. The estimation procedure is adopted from the work of Gospodinov et al. (2005) and steps are herein presented:

Step 1. Given a discrete time stationary parent process $X$ of length $N$, we take the $w - averaged$ process as:
$$X_j^{(w)} = \frac{1}{w}\sum_{i=(j-1)w+1}^{jw} X_i \qquad (12)$$

where $j = 1, 2, \ldots, \frac{N}{w}$ and $w = 1, 2, \ldots, \frac{N}{2}$

Step 2. Take the variance of the aggregated time series as:
$$Var[X^{(w)}] = \frac{1}{(N/w)}\sum_{j=1}^{\frac{N}{w}}(X_j^{(w)} - \bar{X})^2 \qquad (13)$$

where $\bar{X} = \sum_{i=1}^{N} X_i$

Step 3. Obtain the value of β by running a simple linear regression with $\log w$ as the independent variable and $\log[Var(X^{(w)})]$ as the dependent variable. The slope of the resulting equation in (14) of an ordinary least squares regression is the –β.
$$\log Var[X^{(w)}] = \log Var[X] - \beta \log(w) \qquad (14)$$

Step 4. Finally, take the estimate of Hurst as $H = 1 - \beta/2$.

Table 3 below shows the estimates of Hurst Exponent and the Mean Square Error obtained using the Variance Time Plot Analysis (VTP). The estimates of Hurst Exponent for $N$=1024 generally fall below 0.50 while the estimates of Hurst Exponent for $N$=128, 256 are 512 are mostly above 0.50 for all λ parameter values. Considering the Mean Square Error, the estimates at $N$=1024 have the smallest MSE for most λ parameter values. Therefore the estimates using VTP are most efficient at large $N$ values.

**Table 3. Hurst Estimates and MSE using Variance Time Plot Analysis**

| λ | N= Iteration | 128 | 256 | 512 | 1024 | 128 | 256 | 512 | 1024 |
|---|---|---|---|---|---|---|---|---|---|
|   |   | Average Hurst Estimates | | | | Mean Square Error | | | |
| 0.1 | 100 | 0.5215 | 0.5047 | 0.5239 | 0.5015 | 0.0223 | 0.0158 | 0.0188 | 0.0159 |
|     | 500 | 0.5112 | 0.5179 | 0.5199 | 0.4954 | 0.0223 | 0.0184 | 0.0178 | 0.0165 |
|     | 1000 | 0.5145 | 0.5174 | 0.5084 | 0.4958 | 0.0224 | 0.0197 | 0.0182 | 0.0162 |
| 0.5 | 100 | 0.5094 | 0.5082 | 0.5444 | 0.4800 | 0.0200 | 0.0158 | 0.0166 | 0.0163 |
|     | 500 | 0.5141 | 0.5114 | 0.5084 | 0.4917 | 0.0208 | 0.0182 | 0.0179 | 0.0167 |
|     | 1000 | 0.5169 | 0.5014 | 0.5099 | 0.5011 | 0.0209 | 0.0205 | 0.0173 | 0.0175 |
| 1.5 | 100 | 0.5390 | 0.5225 | 0.5051 | 0.4899 | 0.0209 | 0.0176 | 0.0182 | 0.0149 |
|     | 500 | 0.5187 | 0.5184 | 0.5113 | 0.4999 | 0.0240 | 0.0188 | 0.0176 | 0.0159 |
|     | 1000 | 0.5146 | 0.5079 | 0.5129 | 0.4958 | 0.0224 | 0.0195 | 0.0172 | 0.0167 |
| 3.0 | 100 | 0.5197 | 0.4981 | 0.5121 | 0.5026 | 0.0202 | 0.0242 | 0.0146 | 0.0205 |
|     | 500 | 0.5103 | 0.5117 | 0.5094 | 0.5034 | 0.0207 | 0.0196 | 0.0176 | 0.0166 |
|     | 1000 | 0.5145 | 0.5099 | 0.5151 | 0.5011 | 0.0208 | 0.0199 | 0.0176 | 0.0162 |
| 5.0 | 100 | 0.5197 | 0.5139 | 0.5005 | 0.4876 | 0.0187 | 0.0141 | 0.0199 | 0.0205 |
|     | 500 | 0.5192 | 0.5000 | 0.5119 | 0.4989 | 0.0210 | 0.0102 | 0.0192 | 0.0153 |
|     | 1000 | 0.5142 | 0.5034 | 0.5130 | 0.4948 | 0.0226 | 0.0184 | 0.0172 | 0.0163 |
| 7.0 | 100 | 0.5205 | 0.5233 | 0.5067 | 0.4957 | 0.0242 | 0.0188 | 0.0168 | 0.0189 |
|     | 500 | 0.5138 | 0.5074 | 0.5015 | 0.4874 | 0.0232 | 0.0193 | 0.0159 | 0.0179 |
|     | 1000 | 0.5145 | 0.5137 | 0.513 | 0.5038 | 0.021 | 0.0183 | 0.0172 | 0.0162 |

In Figure 3, it can be seen that the estimates are closer to 0.5 when the iterations is done 500 times compared to when the iterations is done 100 times. Therefore, as the number of iterations in the simulation process is increased, the more improved is the resulting estimates. This observation is consistent if we look at the graph of the estimates when the iterations are at 1000 times.

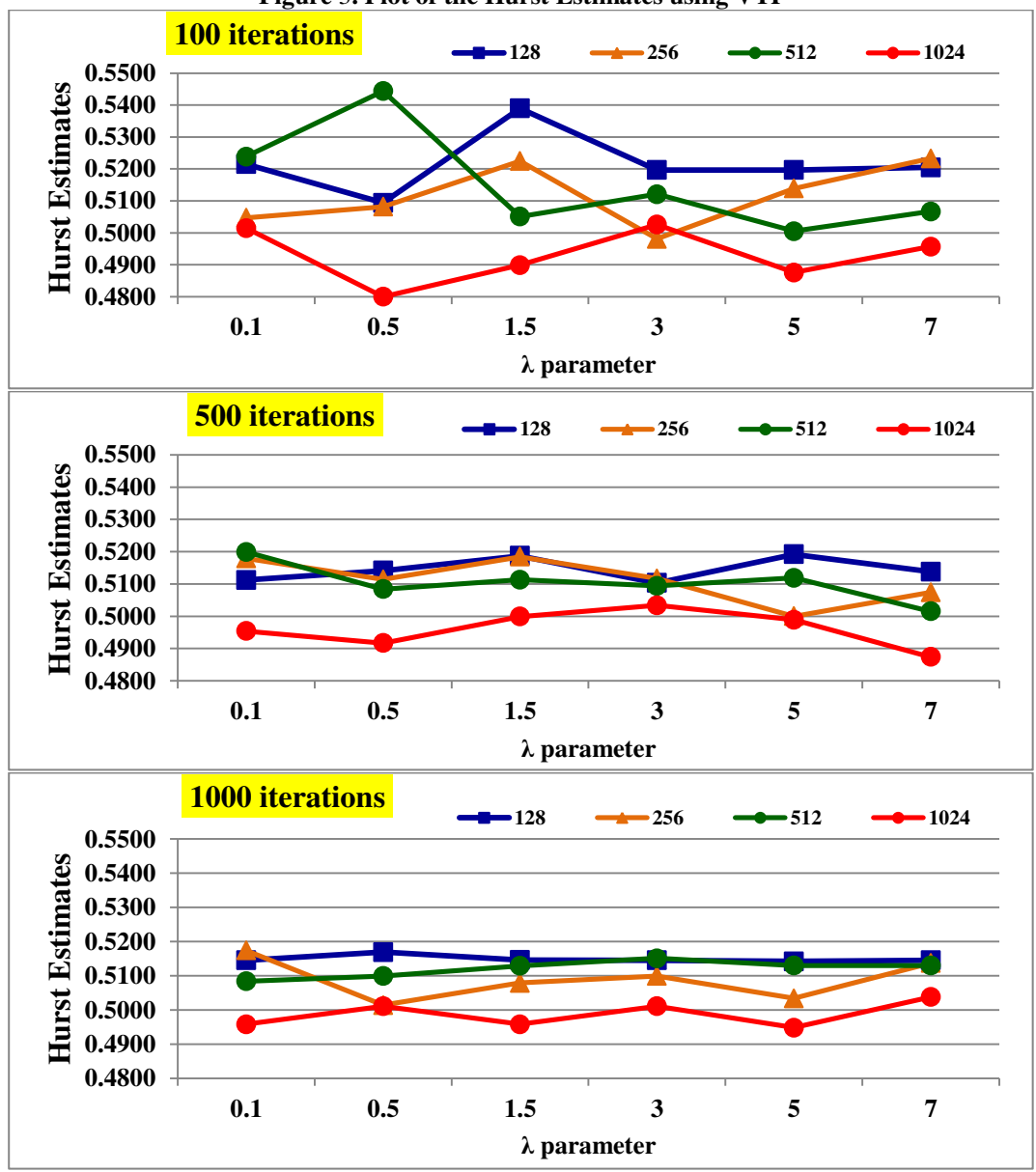

**Figure** 3**. Plot of the Hurst Estimates using VTP**

### 3 Comparison of estimators

In order to test the efficiency of the three methods, we performed computer simulations. We generated samples of data from an exponential distribution of length $L = 2^N$, where N = 7, 8, 9 and 10 i.e. L = 128, 256, 512 and 1024. For each L, we applied the three estimation procedures – Adjusted Rescaled Range Analysis, Detrended Fluctuation Analysis and Variance Time Plot. We repeated the estimation process for 100, 500 and 1000 iteration times and compared the result using the Mean Square Error.

Table 4 below shows the values of the estimates of Hurst Exponent using the three different methods when the sample size $N$ is 128.

The estimation process is repeated for 100, 500 and 1000 times for each method and for each $\lambda$ parameter value. The estimates using the Variance Time Plot Analysis have the highest Mean Square Error (MSE) values while the estimates using the Adjusted Rescaled Range (RSal) Analysis have the lowest Mean Square Error (MSE) values, ranging from 0.0012 to 0.0017 for all $\lambda$ parameter values. Considering the estimates of Hurst Exponent, the Detrended Fluctuation Analysis method has the highest values, ranging from 0.5916 to 0.6079. The Adjusted Rescaled Range method has estimates that are closest to 0.50 in all $\lambda$ parameter values ranging from 0.4964 to 0.5057, while the Variance Time Plot (VTP) Analysis has estimates ranging from 0.4800 to 0.5226. Therefore, the Adjusted Rescaled Range (R/Sal) Analysis is the most efficient method in the estimation of Hurst Exponent when $N$=128.

**Table 4. Comparison when $N$=128 with 100, 500 and 1000 iterations**

| λ | Iterations | Adjusted R/Sal | | DFA | | VTP | |
|---|---|---|---|---|---|---|---|
| | | Hurst | MSE | Hurst | MSE | Hurst | MSE |
| 0.1 | 100 | 0.4964 | 0.0014 | 0.5916 | 0.0152 | 0.5015 | 0.0223 |
| | 500 | 0.5003 | 0.0012 | 0.6024 | 0.0161 | 0.4954 | 0.0223 |
| | 1000 | 0.4986 | 0.0015 | 0.5957 | 0.0155 | 0.4958 | 0.0224 |
| 0.5 | 100 | 0.5057 | 0.0012 | 0.6063 | 0.0172 | 0.4800 | 0.0200 |
| | 500 | 0.4984 | 0.0014 | 0.5954 | 0.0155 | 0.4917 | 0.0208 |
| | 1000 | 0.4999 | 0.0013 | 0.5997 | 0.0162 | 0.5011 | 0.0209 |
| 1.5 | 100 | 0.5013 | 0.0013 | 0.6009 | 0.0162 | 0.4899 | 0.0209 |
| | 500 | 0.4988 | 0.0015 | 0.5959 | 0.0155 | 0.4999 | 0.0240 |
| | 1000 | 0.4986 | 0.0015 | 0.5957 | 0.0155 | 0.4958 | 0.0224 |
| 3.0 | 100 | 0.5020 | 0.0014 | 0.6079 | 0.0172 | 0.5226 | 0.0202 |
| | 500 | 0.4989 | 0.0013 | 0.6013 | 0.0163 | 0.5034 | 0.0207 |
| | 1000 | 0.4998 | 0.0013 | 0.5997 | 0.0162 | 0.5011 | 0.0208 |
| 5.0 | 100 | 0.5016 | 0.0017 | 0.6046 | 0.0180 | 0.4876 | 0.0187 |
| | 500 | 0.5007 | 0.0013 | 0.5980 | 0.0161 | 0.4989 | 0.0210 |
| | 1000 | 0.5027 | 0.0014 | 0.6053 | 0.0174 | 0.4948 | 0.0226 |
| 7.0 | 100 | 0.4965 | 0.0017 | 0.5980 | 0.0166 | 0.4957 | 0.0242 |
| | 500 | 0.5030 | 0.0015 | 0.6075 | 0.0181 | 0.4874 | 0.0232 |
| | 1000 | 0.5012 | 0.0014 | 0.6038 | 0.0174 | 0.5038 | 0.0210 |

Presented in Table 5 are the results of the estimation of the values of Hurst Exponent using three different methods when $N$=256 with 100, 500 and 1000 iterations. The values of Hurst Exponent using DFA are still above 0.50 in all λ parameter values, although there is a slight decrease in the values as $N$ is increased from 128 to 256. The resulting estimates using VTP has a lower MSE compared to the resulting estimates using DFA. The Adjusted Rescaled Range Analysis is the method that produced the estimates closest to 0.50, ranging from 0.4973 to 0.5020. It is also the method that has the lowest Mean Square Errors in all λ parameter values. Hence, the Adjusted Rescaled Range (R/Sal) Analysis is the method that is most efficient when $N$ is 256.

**Table 5. Comparison when $N$=256 with 100, 500 and 1000 iterations**

| λ | Iterations | Adjusted R/Sal | | DFA | | VTP | |
|---|---|---|---|---|---|---|---|
| | | Hurst | MSE | Hurst | MSE | Hurst | MSE |
| 0.1 | 100 | 0.4994 | 0.0008 | 0.5817 | 0.0097 | 0.5047 | 0.0158 |
| | 500 | 0.4999 | 0.0010 | 0.5761 | 0.0091 | 0.5179 | 0.0184 |
| | 1000 | 0.4992 | 0.0009 | 0.5761 | 0.0091 | 0.5174 | 0.0197 |
| 0.5 | 100 | 0.5020 | 0.0009 | 0.5711 | 0.0075 | 0.5082 | 0.0158 |
| | 500 | 0.4994 | 0.0010 | 0.5798 | 0.0098 | 0.5114 | 0.0182 |
| | 1000 | 0.4992 | 0.0009 | 0.5772 | 0.0094 | 0.5014 | 0.0205 |
| 1.5 | 100 | 0.5020 | 0.0009 | 0.5828 | 0.0105 | 0.5225 | 0.0176 |
| | 500 | 0.4983 | 0.0010 | 0.5767 | 0.0093 | 0.5184 | 0.0188 |
| | 1000 | 0.4982 | 0.0009 | 0.5741 | 0.0087 | 0.5079 | 0.0195 |
| 3.0 | 100 | 0.4973 | 0.0009 | 0.5757 | 0.0089 | 0.4981 | 0.0242 |
| | 500 | 0.5005 | 0.0009 | 0.5760 | 0.0092 | 0.5117 | 0.0196 |
| | 1000 | 0.5000 | 0.0010 | 0.5764 | 0.0092 | 0.5099 | 0.0199 |
| 5.0 | 100 | 0.5017 | 0.0009 | 0.5813 | 0.0101 | 0.5139 | 0.0141 |
| | 500 | 0.4974 | 0.0009 | 0.5777 | 0.0092 | 0.5000 | 0.0102 |
| | 1000 | 0.4986 | 0.0009 | 0.5779 | 0.0093 | 0.5034 | 0.0184 |
| 7.0 | 100 | 0.5001 | 0.0009 | 0.5746 | 0.0092 | 0.5233 | 0.0188 |
| | 500 | 0.4974 | 0.0009 | 0.5711 | 0.0083 | 0.5074 | 0.0193 |
| | 1000 | 0.5007 | 0.0009 | 0.5792 | 0.0095 | 0.5137 | 0.0183 |

Table 6. Comparison when *N*=512 with 100, 500 and 1000 iterations

| $\lambda$ | Iterations | Adjusted R/Sal | | DFA | | VTP | |
|---|---|---|---|---|---|---|---|
| | | Hurst | MSE | Hurst | MSE | Hurst | MSE |
| 0.1 | 100 | 0.4982 | 0.0006 | 0.5559 | 0.0050 | 0.5239 | 0.0188 |
| | 500 | 0.4985 | 0.0006 | 0.5590 | 0.0054 | 0.5199 | 0.0178 |
| | 1000 | 0.4991 | 0.0006 | 0.5597 | 0.0054 | 0.5084 | 0.0182 |
| 0.5 | 100 | 0.5020 | 0.0006 | 0.5681 | 0.0067 | 0.5444 | 0.0166 |
| | 500 | 0.5005 | 0.0007 | 0.5621 | 0.0058 | 0.5084 | 0.0179 |
| | 1000 | 0.4987 | 0.0006 | 0.5588 | 0.0054 | 0.5099 | 0.0173 |
| 1.5 | 100 | 0.5025 | 0.0005 | 0.5603 | 0.0053 | 0.5051 | 0.0182 |
| | 500 | 0.4982 | 0.0007 | 0.5562 | 0.0052 | 0.5113 | 0.0176 |
| | 1000 | 0.4986 | 0.0007 | 0.5607 | 0.0058 | 0.5129 | 0.0172 |
| 3.0 | 100 | 0.4969 | 0.0005 | 0.5552 | 0.0047 | 0.5121 | 0.0146 |
| | 500 | 0.4989 | 0.0007 | 0.5596 | 0.0055 | 0.5094 | 0.0176 |
| | 1000 | 0.4988 | 0.0007 | 0.5588 | 0.0053 | 0.5151 | 0.0176 |
| 5.0 | 100 | 0.4983 | 0.0008 | 0.5590 | 0.0056 | 0.5005 | 0.0199 |
| | 500 | 0.4990 | 0.0006 | 0.5575 | 0.0052 | 0.5119 | 0.0192 |
| | 1000 | 0.4993 | 0.0006 | 0.5603 | 0.0056 | 0.5130 | 0.0172 |
| 7.0 | 100 | 0.4959 | 0.0006 | 0.5534 | 0.0044 | 0.5067 | 0.0168 |
| | 500 | 0.5007 | 0.0007 | 0.5630 | 0.0059 | 0.5015 | 0.0159 |
| | 1000 | 0.4993 | 0.0006 | 0.5603 | 0.0056 | 0.5130 | 0.0172 |

Upon examination of the values presented in Table 6, the estimates of Hurst Exponent when *N* is 512 with 100, 500 and 1000 iterations have the smallest Mean Square Error when the method used is Adjusted Rescaled Range (R/Sal) Analysis. The estimates produced using Variance Time Plot (VTP) Analysis is closer to 0.50 compared to the estimates produced using Detrended Fluctuation Analysis (DFA). However, the Mean Square Error (MSE) values of the estimates using Variance Time Plot (VTP) Analysis are bigger in all $\lambda$ parameter values compared to the Mean Square Error (MSE) values of the estimates using Detrended Fluctuation Analysis.

Further observations for the estimates of Hurst Exponent when *N*=1024 with 100, 500 and 1000 iterations are summarized in Table 7. There are major improvements in the estimates produced using Detrended Fluctuation Analysis (DFA) and Variance Time Plot Analysis (VTP) as the iterations used is 1000. However, the achieved improvements of these methods in producing efficient estimates of Hurst Exponent are not enough to make them more efficient than the Adjusted Rescaled Range Analysis. The Adjusted Rescaled Range Analysis (R/Sal) still has the lowest Mean Square Errors in all $\lambda$ parameter values when *N* is 1024.

Table 7. Comparison when N=1024 with 100, 500 and 1000 iterations

| $\lambda$ | Iterations | Adjusted R/Sal | | DFA | | VTP | |
|---|---|---|---|---|---|---|---|
| | | Hurst | MSE | Hurst | MSE | Hurst | MSE |
| 0.1 | 100 | 0.4963 | 0.0005 | 0.5427 | 0.0031 | 0.5215 | 0.0159 |
| | 500 | 0.4994 | 0.0005 | 0.5474 | 0.0035 | 0.5112 | 0.0165 |
| | 1000 | 0.4994 | 0.0005 | 0.5478 | 0.0035 | 0.5145 | 0.0162 |
| 0.5 | 100 | 0.4989 | 0.0004 | 0.5481 | 0.0034 | 0.5094 | 0.0163 |
| | 500 | 0.4984 | 0.0005 | 0.547 | 0.0035 | 0.5141 | 0.0167 |
| | 1000 | 0.4988 | 0.0005 | 0.5473 | 0.0036 | 0.5169 | 0.0175 |
| 1.5 | 100 | 0.4994 | 0.0005 | 0.5486 | 0.0035 | 0.5390 | 0.0149 |
| | 500 | 0.4991 | 0.0005 | 0.5474 | 0.0034 | 0.5187 | 0.0159 |
| | 1000 | 0.4984 | 0.0005 | 0.5469 | 0.0035 | 0.5146 | 0.0167 |
| 3.0 | 100 | 0.5003 | 0.0005 | 0.5457 | 0.0034 | 0.5197 | 0.0205 |
| | 500 | 0.4996 | 0.0004 | 0.5482 | 0.0035 | 0.5103 | 0.0166 |
| | 1000 | 0.4994 | 0.0005 | 0.5478 | 0.0035 | 0.5145 | 0.0162 |
| 5.0 | 100 | 0.4982 | 0.0005 | 0.5455 | 0.0034 | 0.5197 | 0.0205 |
| | 500 | 0.4999 | 0.0005 | 0.5485 | 0.0037 | 0.5192 | 0.0153 |
| | 1000 | 0.4995 | 0.0005 | 0.5479 | 0.0035 | 0.5142 | 0.0163 |
|  | 100 | 0.4953 | 0.0004 | 0.5419 | 0.0028 | 0.5205 | 0.0189 |
| 7.0 | 500 | 0.4990 | 0.0005 | 0.5474 | 0.0036 | 0.5138 | 0.0179 |
|  | 1000 | 0.4994 | 0.0005 | 0.5478 | 0.0035 | 0.5145 | 0.0162 |

## Acknowledgements

We would like to acknowledge the thesis review committee: Dr. Eveyth P. Deligero, chairman and Mr. Ronald D. Estrada and Mr. Anthony F. Capili, members.